\documentstyle[12pt,epsf,epsfig,cite]{article}

\def\beq{\begin{equation}}
\def\eeq{\end{equation}}

\def\as{\alpha_{\mbox{\scriptsize s}}}
\def\eqref#1{(\ref{#1})}
\def\qcold{\hat{q}_{\mbox{\scriptsize{cold}}}}
\def\qhot{\hat{q}_{\mbox{\scriptsize{hot}}}}
\def\EHQ{E_{\mbox{\scriptsize{HQ}}}}
\def\cO#1{{\cal{O}}\left(#1\right)}
\def\GeV{{\rm Ge\!V}}
\def\MeV{{\rm Me\!V}}
\def\fm{{\rm fm}}

\def\la{\mathrel{\mathpalette\fun <}}

\def\fun#1#2{\lower3.6pt\vbox{\baselineskip0pt\lineskip.9pt
  \ialign{$\mathsurround=0pt#1\hfil##\hfil$\crcr#2\crcr\sim\crcr}}}
 \newskip\humongous \humongous=0pt plus 1000pt minus 1000pt
 \def\caja{\mathsurround=0pt} \def\eqalign#1{\,\vcenter{\openup1\jot
 \caja   \ialign{\strut \hfil$\displaystyle{##}$&$
 \displaystyle{{}##}$\hfil\crcr#1\crcr}}\,} \newif\ifdtup

\begin{document}
\begin{flushright}
LPT-Orsay-01/58\\
{BNL-NT--01/9} \\
\end{flushright}
\vskip1.5cm
\begin{center}
{\Large\bf Heavy Quark Colorimetry of QCD Matter}
\end{center}
\medskip
\begin{center}
{\large Yu.L. Dokshitzer$^{a}$ and D.E. Kharzeev$^{a,b}$}
\end{center}
\vskip1cm
\begin{center}
$^{a)}$ LPTHE, Universit{\'e} Pierre et Marie Curie (Paris VI),\\
4, Place Jussieu, 75252 Paris, \\
and \\
LPT, Universit\'e Paris Sud, 91405 Orsay,
France
\vskip0.3cm
$^{b)}$ Physics Department,\\
Brookhaven National Laboratory,\\
 Upton, New York 11973-5000, USA\footnote{Permanent address}
\end{center}

\vskip0.3cm

%\preprint{BNL-NT/01-XXX\\
%     hep-ph/0106202 \\
%     May 2001}
%
%\title{Heavy Quark Colorimetry of QCD Matter}
%
%\author{Yu.L.\ Dokshitzer\footnote{on leave from 
%St.\ Petersburg Nuclear Institute, Gatchina, St.~Petersburg 188350, Russia}
% and D.E.\ Kharzeev\footnote{permanent address: Physics Department,
%Brookhaven National Laboratory, 
% Upton, New York 11973-5000, USA}\\
%LPTHE, Universit{\'e} Pierre et Marie Curie (Paris VI),
%4, Place Jussieu, \\ 75252 Paris, 
%and
% LPT, Universit\'e Paris Sud, 91405 Orsay, France
%}

\begin{abstract}
We consider propagation of heavy quarks in QCD matter. Because of
large quark mass, the radiative quark energy loss appears to be
qualitatively different from that of light quarks at all energies of
practical importance.  Finite quark mass effects lead to an in-medium
enhancement of the heavy--to--light $D/\pi$ ratio at moderately large
(5--10 $\GeV$) transverse momenta. For hot QCD matter a
% spectacularly
large enhancement is expected, whose magnitude and shape are
exponentially sensitive to the density of colour charges in the
medium.
\end{abstract}

\section{Introduction}

The energy loss of heavy quarks in QCD matter has excited considerable
interest. In particular, collisional energy loss in quark--gluon
plasma has been evaluated \cite{elastic} by the methods of
finite--temperature QCD (see \cite{jp} for a comprehensive
review). However, the energy loss of fast partons in medium is
dominated by radiation of gluons \cite{GW,BDPS,Zakh} and, to the best of
our knowledge, a consistent treatment of radiative energy loss of
heavy quarks in QCD matter is still lacking. Practical importance of
this issue has become clear from a number of studies
\cite{phen} which showed a strong dependence of charmed hadron 
and lepton spectra on the assumed heavy quark energy loss.

In this letter we follow the recent analysis of the quenching of
inclusive particle spectra in QCD medium~\cite{BDMSquen} to compare
the yields of hadrons from light and heavy quarks produced in hard
interactions in heavy ion collisions.  We find that the specific
pattern of gluon bremsstrahlung off heavy quarks (the dead cone
effect) results in a much smaller heavy quark quenching, leading to 
a relative enhancement of heavy particle production in heavy ion
collisions.

\section{Medium induced gluon radiation -- a brief summary}

We first recall the basic features of gluon radiation caused by
propagation of a fast parton (quark) through QCD medium.

As was pointed out in \cite{BDPS}, the accompanying radiation is
determined by multiple rescattering of the radiated gluon in the
medium.  The gluon, during its formation time
\beq 
   t_{form} \simeq \frac{\omega}{k_{\perp}^2}\,,
\label{form}
\eeq
accumulates a typical transverse momentum
\beq 
   k_{\perp}^2 \simeq \mu^2 \ {t_{form} \over \lambda}, \label{walk}
\eeq
with $\lambda$ the 
%%% gluon 
mean free path and $\mu^2$ the characteristic momentum transfer
squared in a single scattering.  This is the random walk pattern with
an average number of scatterings given by the ratio
$t_{form}/\lambda$.

Combining (\ref{walk}) and (\ref{form}) we obtain
\beq\label{Ncoh}
  N_{coh}=\frac{t_{form}}{\lambda} = \sqrt{\frac{\omega}{\mu^2\,\lambda}}
\eeq
describing the number of scattering centres which participate, {\em
coherently}, in the emission of the gluon with a given energy
$\omega$.  For sufficiently large gluon energies,
$\omega>\mu^2\lambda$, when the coherent length exceeds the mean free
path, $N_{coh}>1$.  In this situation the standard Bethe-Heitler
energy spectrum per unit length describing {\em independent}\/
emission of gluons at each centre gets suppressed:
\beq\label{spec}
 \frac{dW}{d\omega dz} = \left(
\frac{dW}{d\omega dz}\right)^{\mbox{\scriptsize BH}} \cdot\frac1{N_{coh}}
= \frac{\as C_R}{\pi\omega\,\lambda}\cdot\sqrt{\frac{\mu^2\,\lambda}{\omega}} 
=  \frac{\as C_R}{\pi\omega} \sqrt{\frac{\hat{q}}{\omega}}.
\eeq
Here $C_R$ is the ``colour charge'' of the parton projectile
($C_R=C_F=\frac{N_c^2 - 1}{2 N_c} = 4/3$ for the quark case we are
interested in).

In \eqref{spec} we have substituted the characteristic ratio 
$\mu^2/\lambda$ by the so-called gluon 
{\em transport coefficient}\/
\cite{BDMPS1}
\beq
 \hat{q} \equiv 
%%%% {\mu^2 \over \lambda} =  
 \rho \ \int {d \sigma \over dq^2}\ q^2\ dq^2, \label{qhat}
\eeq
which is proportional to the density $\rho$ of the scattering centres
%color charges 
in the medium and describes the typical momentum transfer in the gluon
scattering off these centres.
%%% 

The transport coefficient 
for cold nuclear matter was expressed in \cite{BDMPS1} as
% through the nuclear density and the gluon structure function of the
% nucleon,
\beq 
\hat{q} \simeq {4 \pi^2 \alpha_s N_c\over N_c^2-1}\ 
\rho\ [xG(x, Q^2)], \label{qhatcold}
\eeq
with $\rho \simeq 0.16\ {\rm fm^{-3}}$  the average nuclear
density and $[xG(x)]$ the gluon density in a nucleon.  The latter
should be evaluated at the characteristic scale $Q^2 \sim \hat{q} L$,
which makes (\ref{qhatcold}), strictly speaking, an equation for
$\hat{q}$. Given a slow logarithmic $Q^2$ dependence of the gluon
density, this equation can be solved iteratively.  Choosing for the
sake of estimate $L \simeq 5$ fm results in $Q \sim 0.5\ {\rm GeV}$.
Taking at this scale $\alpha_s \simeq 0.5$ and $[xG(x)] \simeq 1$ (at
$x < 0.1$),  yields 
%
% $\hat{q}_{cold} \simeq 0.005\ {\rm GeV^3}$.
\beq
 \qcold \simeq 0.01 \ {\rm GeV^3} \simeq 8 \ \rho. \label{numqhat}
\eeq  
This value 
%% may be compared to 
is an agreement with 
the result of the analysis of the gluon $p_{\perp}$ broadening from
the experimental data on $J/\psi$ transverse momentum distributions
\cite{KNS}, which in the present notation yielded
\beq 
\hat{q} = (9.4 \pm 0.7) \> \rho\,.
\eeq
An estimate \cite{BDMPS1} for a hot medium based on perturbative
treatment of gluon scattering in quark--gluon plasma with $T\sim
250$~MeV resulted in the value of the gluon transport coefficient of
about factor {\em twenty}\/ larger than \eqref{numqhat}:
%
%In what follows, to be on a conservative side, we shall accept for the
%hot medium a somewhat smaller value 
\beq
   \qhot \simeq 0.2\ \rm{GeV}^3\>\simeq\> 20\, \qcold \,. 
\label{qhathot}
\eeq

Multiplying \eqref{spec} by the length $L$ of the medium
traversed,\footnote{For the sake of simplicity we assume here that the
medium is static and uniform.}  we arrive at the following expression
for the inclusive energy distribution of gluons radiated by a quark:
\beq\label{omega1}
  \frac{dW}{d\omega } \simeq \frac{\as C_F}{\pi\,\omega}
  \sqrt{\frac{\omega_1}{\omega}}, \qquad \omega \><\> 
  \omega_1\equiv  \hat{q}L^2\,.
\eeq 
A more accurate treatment shows that the semi-quantitative estimate
\eqref{spec} is valid for $\omega\la \omega_1/2$, while at very large
gluon energies $\omega\gg\omega_1$ the gluon yield is very small and
decreases as $dI/d\omega\propto \omega^{-3}$ 
%%%
(for details, see \cite{BDMPS1}).

The fact that the medium induced radiation vanishes for
$\omega>\omega_1$ has a simple physical explanation, as according to
\eqref{Ncoh} the formation time of such gluons starts to exceed the
length of the medium:
\[
 t_{form}= \lambda\cdot\sqrt{\frac{\omega}{\mu^2\lambda}} 
= \sqrt{\frac{\omega}{\hat{q}}} = L\cdot
\sqrt{\frac{\omega}{\omega_1}} \>>\> L\,.
\] 

Another important feature of medium induced radiation is the relation
between the transverse momentum and the energy of the emitted gluon.
Indeed, from \eqref{form} and \eqref{walk} (see also \eqref{qhat}) we
derive
\beq
  k_{\perp}^2 \simeq \sqrt{\hat{q}\ \omega} \label{star}.
\eeq 
This means that the angular distribution of gluons with a given energy
$\omega$ is concentrated at a characteristic energy- (and medium-)
dependent emission angle
\beq\label{angle}
\theta \simeq \frac{k_\perp}{\omega} 
\sim \left(\frac{\hat{q}}{\omega^3}\right)^{1/4}. 
\eeq

\section{Radiation off heavy quarks: dead cone}

Gluon bremsstrahlung off a heavy quark differs from the case of a
massless parton (produced in a process with the same hardness scale)
in one respect: gluon radiation is suppressed at angles smaller than
the ratio of the quark mass $M$ to its energy $E$.  
Indeed, the distribution of soft gluons radiated by a heavy quark is
given by
\beq
dP  = {\as\ C_F \over \pi}\ 
{d\omega \over \omega}\ {k_{\perp}^2 \,dk_\perp^2\over 
(k_{\perp}^2 + \omega^2 \theta_0^2)^2}, \qquad
\theta_0\equiv\frac{M}{E}\,,
\label{dist}
\eeq
where 
%$z$ is the fraction of the quark's momentum carried away by the
% gluon with the transverse momentum $ k_{\perp}$, and 
the strong coupling constant $\alpha_s$ should be evaluated at the scale
determined by the denominator of (\ref{dist}).  Equating, in the
small-angle approximation, $k_\perp$ with $\omega\theta$ we
conclude that the formula (\ref{dist}) differs from the standard
bremsstrahlung spectrum
\beq
 dP_0\> \simeq \> \frac{\as\,C_F}{\pi}
 \frac{d\omega}{\omega}\,\frac{dk_\perp^2}{k_\perp^2}
\> =\> \frac{\as\,C_F}{\pi} \frac{d\omega}{\omega}\,\frac{d\theta^2}{\theta^2}
\eeq
by the factor 
\beq\label{factor}
 dP_{\mbox{\scriptsize HQ}} = dP_0\cdot \left( 
1+\frac{\theta_0^2}{\theta^2}\right)^{-2}
\eeq

This effect is known as the ``dead cone'' phenomenon.  Suppression of
small-angle radiation has a number of interesting implications, such
as perturbative calculability of (and non-perturbative $\Lambda/M$
corrections to) heavy quark fragmentation functions~\cite{DKT,NW},
multiplicity and energy spectra of light particles accompanying hard
production of a heavy quark~\cite{hqmulspec}.

\begin{figure}[h]
\begin{minipage}{12cm}
\begin{center}
\epsfig{file=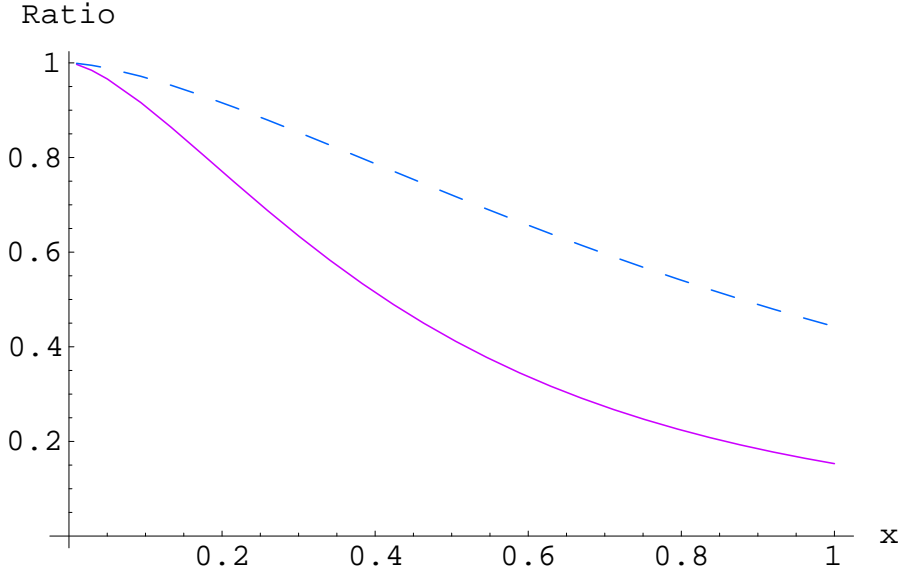, width=12cm}
\end{center}
\end{minipage}
\begin{minipage}{14cm}
\caption{
Ratio of gluon emission spectra off charm and light quarks
for quark momenta $p_\perp=10\,\GeV$ (solid line)
and $p_\perp=100\,\GeV$ (dashed); $x=\omega/p_\perp$ .}
\label{supprfig}
\end{minipage}
\end{figure}

In the present context we should compare the angular distribution
of gluons induced by the quark propagation in the medium with the
size of the dead cone.
To this end, for the sake of a semi-quantitative estimate, we 
substitute the characteristic angle \eqref{angle} into 
the dead cone suppression factor \eqref{factor} and combine it with
the radiation spectrum \eqref{spec} to arrive at
\beq
I(\omega) = \omega {d W \over d \omega} = {\alpha_s\ C_F \over \pi}
\sqrt{{\omega_1 \over 
\omega}} \ {1 \over (1 + (\ell\, \omega)^{3/2})^2}, 
\label{eq:spechq}
\eeq 
where
\beq
  \ell \equiv  \hat{q}^{-1/3}\ \left({M \over E}\right)^{4/3}. \label{apar}
\eeq
The suppression factor \eqref{eq:spechq} is displayed in
Fig.~\ref{supprfig} for two quark energies ($x=\omega/p_\perp$).

To see whether the finite quark mass essentially affects the medium
induced gluon yield, we need to estimate the product $\ell\omega$ for
the maximal gluon energy $\omega\simeq\omega_1$ to which the original
distribution
\eqref{spec} extends:
\beq
 \ell\omega_1= \hat{q}^{-1/3}\ \left({M \over E}\right)^{4/3}\cdot
\hat{q}L^2
= \left(\frac{\EHQ}{E}\right)^{4/3}\,, \qquad \EHQ\equiv M\sqrt{\hat{q}L^3}.
\eeq
This shows that the quark mass becomes irrelevant when the quark
energy exceeds the characteristic value $\EHQ$ which depends on the
size of the medium and on its ``scattering power'' embodied into the
value of the transport coefficient.

Which regime is realized in the experiments on heavy quark 
production in nuclear collisions? 
Taking $M = 1.5\ \rm{GeV}$ for charm quarks and using the values
\eqref{numqhat} and \eqref{qhathot} 
we estimate
\begin{eqnarray}
\label{EHQc}
\EHQ &=& \sqrt{\qcold}\ L^{3/2}\ M \simeq \ 20\ {\GeV}
\left({L \over 5\, {\rm fm}}\right)^{3/2}, \\
\label{EHQh}
\EHQ &=& \sqrt{\qhot}\ L^{3/2}\ M \simeq \ 
%%% 65
92\  {\GeV} 
\left({L \over 5\, {\rm fm}}\right)^{3/2},
\end{eqnarray}
for the cold and hot matter, respectively.  We observe that for the
transverse momentum (energy) distributions of heavy mesons the
inequality $E \ll \EHQ$ always holds in practice, especially for the
hot medium.  We thus conclude that the pattern of medium induced gluon
radiation appears to be {\em qualitatively different for heavy and
light quarks}\/ in the kinematical region of practical interest.

\section{Quenching}

The issue of in-medium quenching of inclusive particle spectra was
recently addressed in \cite{BDMSquen}.  The $p_{\perp}$ spectrum is
given by the convolution of the transverse momentum distribution in an
elementary hadron--hadron collision, 
%%% $d \sigma_{pp}/d p_{\perp}^2$, 
evaluated at a shifted value $p_\perp+\epsilon$, with the 
distribution $D(\epsilon)$ in the energy $\epsilon$ lost 
by the quark to the medium-induced gluon radiation:
\beq
{d \sigma^{med} \over d p_{\perp}^2} = \int d \epsilon \ D(\epsilon)\ 
{d \sigma^{vac} \over d p_{\perp}^2}( p_{\perp} + \epsilon )
\equiv {d \sigma^{vac} \over d p_{\perp}^2}( p_{\perp}) \cdot Q(p_\perp),
  \label{defpt}
\eeq
with $Q(p_\perp)$ the medium dependent {\em quenching factor}.  The
two facts, namely that in the essential region $\epsilon\ll p_\perp$
and that the vacuum cross section is a steeply falling function, allow
one to simplify the calculation of the quenching factor $Q$ by
adopting the exponential approximation for the $\epsilon$-integral in
\eqref{defpt}:
\beq\label{Qfint}
 Q(p_\perp) \simeq \int d \epsilon \ D(\epsilon)\ \exp\left\{ 
\frac{\epsilon}{p_\perp} \cdot {\cal{L}} \right\} \,, \qquad 
{\cal{L}} \equiv \frac{d}{d\ln p_\perp} 
\ln \left[ {d \sigma^{vac} \over d p_{\perp}^2}( p_{\perp})\right].
\eeq 
This integral results in the {\em Mellin moment}\/ of the quark
distribution,  
\beq\label{Qexp}
Q(p_\perp)\>=\> 
\tilde{D}(\nu) = \exp\left[ - \nu\ \int_0^{\infty} d \omega \ 
 N(\omega)\ e^{-\nu \omega}\right], \quad \nu=\frac{{\cal{L}}}{p_\perp},
\label{tildf}
\eeq  
where $N(\omega)$ is the {\em integrated gluon multiplicity}\/ defined
according to (see \cite{BDMSquen} for details) 
\beq
N(\omega) \equiv \int_{\omega}^{\infty} d\omega' \, {d W (\omega') \over d
\omega'}\,.
 \label{intm}
\eeq
Here we present an approximate evaluation based on a simplified energy
spectrum (\ref{eq:spechq}). Integration of (\ref{eq:spechq}) leads to the
following expression for the gluon multiplicity (\ref{intm}):
\beq
N(\omega) = {2 \alpha_s\ C_F \over \pi}\ \sqrt{\ell\omega_1}\ K(\ell\omega),
\eeq
where 
\beq
\eqalign{
K(x) &\equiv {1 \over \sqrt{x}} 
- {2 \pi \over 3 \sqrt{3}} + {x \over 3
\left(1 + x^{3/2}\right)} +  {4\over 3 \sqrt{3}}\ 
\arctan\left({-1 + 2 \sqrt{x} \over \sqrt{3}}\right)  \cr
& +   {2 \over 9}\ \ln\left[{1 - \sqrt{x} 
+ x \over (1 + \sqrt{x} )^2} \right]. 
\label{exacmult}
}
\eeq   
The function $K(x)$ has essential support in the region $\ell\omega=x
\ll 1$, that is where the gluon yield is not suppressed by the quark
mass (dead cone) effect (cf.\ \eqref{eq:spechq}).  To find the quenching
factor we need to evaluate its Mellin transform according to
(\ref{Qexp}), with $\nu$
% = n p_{\perp}/(a^2 + p_{\perp}^2)$ 
given in (\ref{ftp}). The essential energies $\omega$ in the
multiplicity function in \eqref{Qexp} are then restricted to
$\omega\la 1/\nu$. It is easy to check that, due to the bias effect
(large value of ${\cal{L}}\sim 10$), this restriction is stronger than
the quark mass energy cutoff:
\beq
{\nu \over \ell} 
%%% = \frac{ n \, p_{\perp}}{a^2+p_\perp^2}\cdot
%%% \hat{q}^{1/3}\left(\frac{E}{M}\right)^{4/3} 
= \frac{{\cal{L}}}{p_\perp} \cdot
 \hat{q}^{1/3}\left(\frac{E}{M}\right)^{4/3}
\simeq  {\cal{L}}\left(\frac{p_\perp\,\hat{q}}{M^4}\right)^{1/3} \>>\> 2
\eeq  
for $p_\perp>5$~GeV, even for cold matter.  Therefore, in this
practically interesting domain, we can approximate
\beq
N(\omega) \simeq  {2 \alpha_s\ C_F \over \pi}\ \sqrt{\omega_1}\ 
\left( {1 \over \sqrt{\omega}} - {8 \pi \sqrt{\ell} \over 9
\sqrt{3}}\right) \left[\, 1 + \cO{\ell\omega}\, \right], \label{approx}
\eeq
and get 
\beq
\tilde{D}(\nu) \simeq \exp\left[ - {2 \alpha_s\ C_F \over \pi}\
\sqrt{\hat{q}} L\  
\left(\sqrt{\pi \nu} -  
{8 \pi \sqrt{\ell} \over 9 \sqrt{3}}\right)\right].
\eeq    
The use of (\ref{Qexp}) furnishes our final result:
\beq
Q_H(p_{\perp}) \simeq \exp \left[- {2 \alpha_s C_F \over \sqrt{\pi}}\ 
%%% \left({n\, p_{\perp} \over a^2 +  p_{\perp}^2}\right)^{1/2} 
L\,\sqrt{\hat{q}\frac{{\cal{L}}_H}{p_\perp}}
%%% \sqrt{\hat{q}} 
 + 
{16 \alpha_s C_F \over 9 \sqrt{3}} L
\left( \frac{ \hat{q}\> \> M^2}{M^2+p_\perp^2}\right)^{1/3}  \right].
\label{finres}
\eeq
The first term in the exponent in (\ref{finres}) represents the
quenching of the transverse momentum spectrum which is universal for
the light and heavy quarks,
\[
Q_L(p_{\perp}) \simeq \exp \left[- {2 \alpha_s C_F \over \sqrt{\pi}}\ 
L\,\sqrt{\hat{q}\frac{{\cal{L}}_L}{p_\perp}}\, \right] 
\]
(modulo the difference of the ${\cal{L}}$ parameters 
determined by the $p_{\perp}$ distributions in the vacuum).  The second
term is specific for heavy quarks. It has a positive sign, which means
that the suppression of the heavy hadron $p_{\perp}$ distributions is
always smaller than that for the light hadrons. This is a
straightforward consequence of the fact that the heavy quark mass
suppresses gluon radiation. At very high transverse momenta, both
terms vanish -- this is in accord with the QCD factorization theorem,
stating that the effects of the medium should disappear as $p_{\perp}
\to \infty$. How fast this regime is approached depends, however, on
the properties of the medium encoded in the value of the transport
coefficient $\hat{q}$ and in the medium size $L$.

Constructing the ratio of the quenching functions, we estimate the
heavy-to-light enhancement factor as
\begin{equation}
\frac{Q_H(p_{\perp})}{Q_L(p_{\perp})} \>\simeq\> 
 \exp \left[ {16 \alpha_s C_F \over 9 \sqrt{3}} L
\left( \frac{ \hat{q}\> \> M^2}{M^2+p_\perp^2}\right)^{1/3}  \right].
\label{eq:ratio}
\end{equation}
This simple expression provides a reasonably good approximation to
more accurate quantitative results presented below.

\section{Quantitative estimates}

To perform a quantitative calculation of the transverse momentum
distributions, our considerations have to be combined with a realistic
treatment of nuclear geometry, including its centrality dependence,
gluon multiple interactions in the initial state (the ``Cronin
effect''), and a model for the time evolution of QCD matter in the
final state.  Such an analysis is beyond the scope of the present
paper.
Nevertheless, we provide some semi-quantitative illustrations of the
expected consequences of our results, based on a simplified model of a
static uniform medium and a fixed quark path length $L$. 

\begin{figure}[h]
\begin{minipage}{12cm}
\begin{center}
\epsfig{file=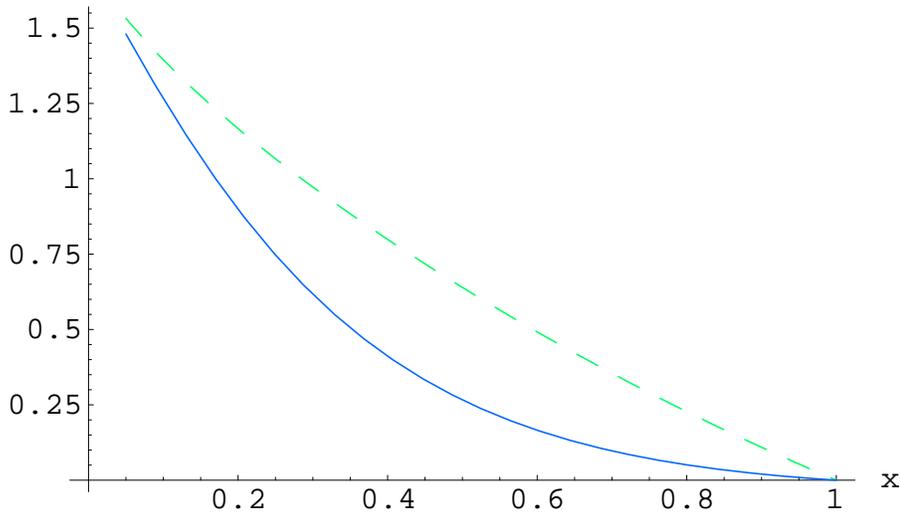, width=12cm}
\end{center}
\end{minipage}
\begin{minipage}{14cm}
\caption{Comparison of energy distributions $\sqrt{x}I(x)$  
of gluons radiated off charm (solid line) and light (dashed line)
quarks in hot matter with $\hat{q}=0.2\,\GeV^3$ ($p_\perp=10\,\GeV$,
$L=5\,\fm$).}
\label{dIdom}
\end{minipage}
\end{figure}

In what follows we use the full inclusive soft gluon emission spectrum
off a massless quark from \cite{BDMPS1}, taking into account the first
non-soft correction $\cO{x}$:
\begin{equation}
  I(x) \>=\> (1-x)\frac{C_F\as}{\pi}
  \ln\left(\cosh^2\left(\sqrt\frac{\omega_1}{4\omega}\right) -
  \sin^2\left(\sqrt\frac{\omega_1}{4\omega}\right) \right), \qquad
  x\equiv \frac{\omega}{E}\,,
\label{eq:exspec}
\end{equation}
which, for the heavy quark case, is supplied with the dead-cone
suppression factor \eqref{factor}.
To illustrate the ``dead cone'' effect, in Fig.~\ref{dIdom} a
comparison of the inclusive one-gluon emission spectra off light and
charm quarks is shown, for hot medium with $L=5\,\fm$.

As we have discussed above, the quenching of $p_{\perp}$ distributions
of heavy hadrons caused by QCD matter is much weaker than for
pions. This is because the gluon cloud around the heavy quark is
``truncated'' by the large quark mass, in a way which depends on the
properties of the medium. This interesting effect can be illustrated
by the transverse momentum dependence of the ratio of hadrons
originating from the fragmentation of heavy and light quarks (for
example, $D/\pi$ ratio) in heavy ion collisions.

To estimate this ratio we need to know the behaviour of the vacuum
spectra.
For {\em light}\/ hadrons (which we assume to originate from light
quarks\footnote{we remark that the quenching of hadrons produced in
the fragmentation of primary {\em gluons}\/ contains an additional
factor $9/4\simeq 2$ in the exponent}) the parameterization of the
$p_{\perp}$ distribution which describes the first RHIC
hadroproduction data for $p_{\perp}$ up to $\sim 6$~GeV~\cite{param}:
\beq
{d \sigma \over d p_{\perp}^2} = A \ \left({1 \over p_0 + 
p_{\perp}}\right)^m,
\eeq
with $p_0 =
%%% 2.74
1.71~\GeV$ and $m = 12.42$.
%%%13.65$. 

The $p_{\perp}$ distributions of $D$ mesons produced in hadron
collisions were experimentally found \cite{E769} to be well
described by the following simple parameterization:
\beq
{d \sigma \over d p_{\perp}^2} = C \ \left({1 \over {b m_c^2 + 
p_{\perp}^2}}\right)^\frac{n}{2},
\eeq
with $b = 1.4 \pm 0.3$, $n = 10.0 \pm 1.2$, and $m_c = 1.5$ GeV. 
This parameterization also provides a good fit to the theoretical 
calculations \cite{Frix97} in perturbative QCD. 
Using this parameterization in \eqref{Qfint} gives 
\beq\label{ftp}
 \nu \>=\> \frac{n\ p_\perp}{a^2+p_\perp^2}\,, \qquad a^2\equiv bm_c^2\,.
\eeq
for the $\nu$ parameter in \eqref{Qexp}.

\begin{figure}[h]
\begin{center}
\epsfig{file=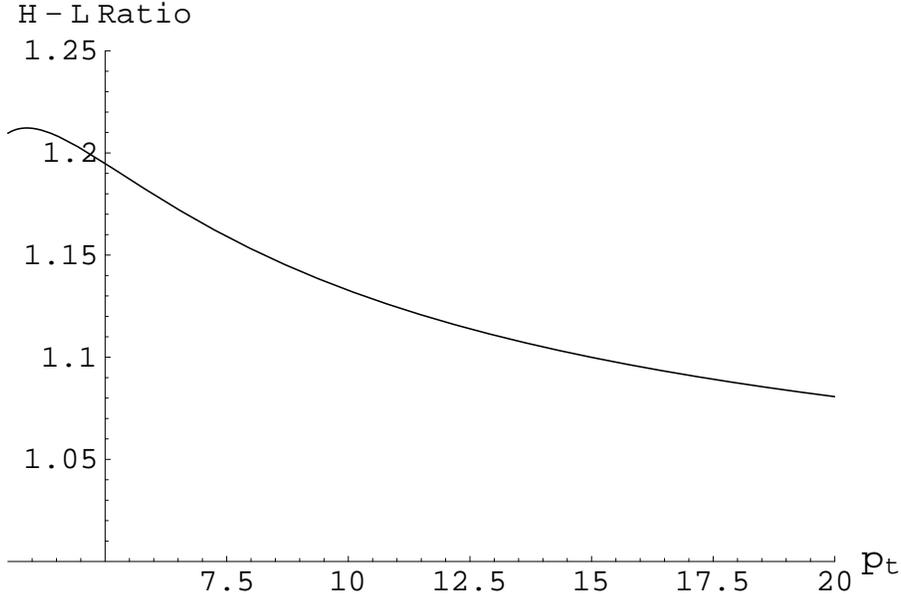, width=12cm}
\end{center}
\caption{The ratio of quenching factors $Q_H(p_{\perp})/Q_L(p_{\perp})$ 
for charm and light quarks in cold nuclear matter
($\hat{q}=0.01\,\GeV^3$, $L=5\,\fm$).
}
\label{Fcold}
\end{figure}

Fig.~\ref{Fcold} shows the ratio of quenching factors (\ref{finres})
for heavy and light quarks in cold nuclear matter ($L=5\,\fm$), which
is relevant for large-$p_\perp$ particle production in $pA$
collisions.  A small value of the transport coefficient, $\qcold\simeq
0.01\,\GeV^3$, translates into a $\sim$ 15\% enhancement of the
heavy-to-light ratio (about 1\%\ for $L=2\,\fm$).

\begin{figure}[h]
\begin{minipage}{12cm}
\begin{center}
\epsfig{file=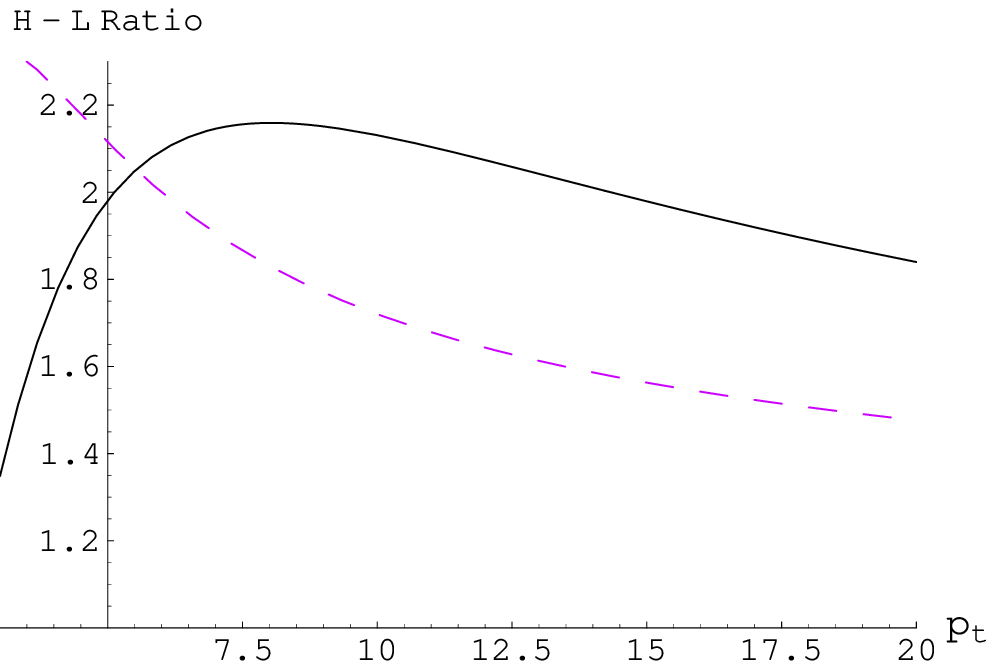, width=11cm}
\end{center}
\end{minipage}

\begin{minipage}{12cm}
\begin{center}
\epsfig{file=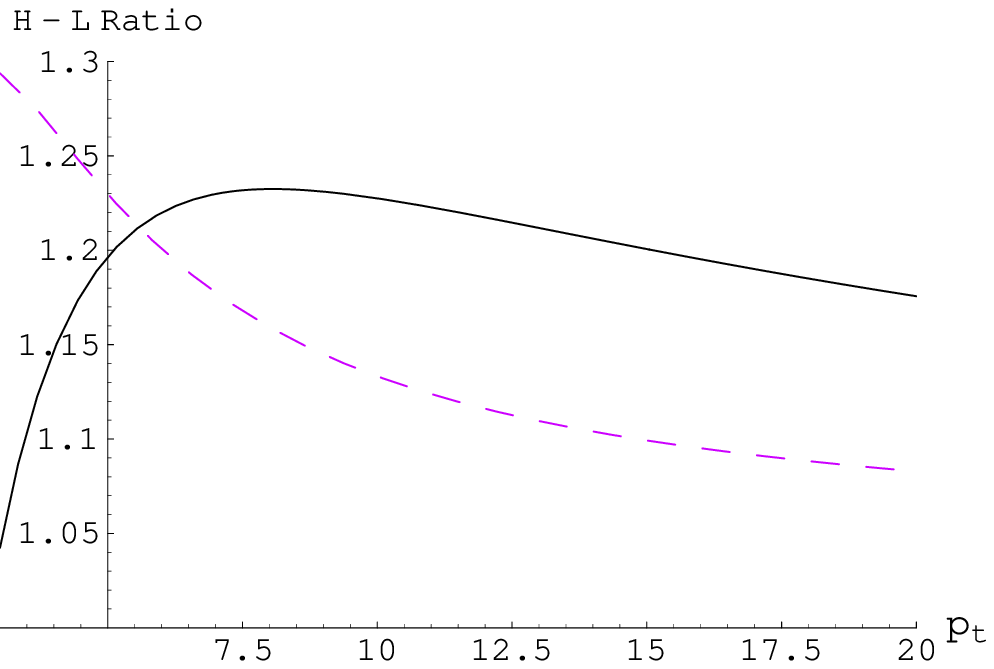, width=11cm}
\end{center}
\end{minipage}
\begin{minipage}{12cm}
\caption{The ratio of quenching factors $Q_H(p_{\perp})/Q_L(p_{\perp})$ 
for charm and light quarks 
in hot matter with $\hat{q}=0.2\,\GeV^3$ ($L=5\,\fm$ upper panel;
$L=2\,\fm$ lower panel).  Solid lines correspond to unrestricted
gluon radiation, while the dashed lines are based on the calculation
with the cut on gluon energies $\omega>0.5\,\GeV$.}
\label{Fhot}
  \end{minipage}
\end{figure}

According to \eqref{eq:ratio}, we expect a larger quenching ratio for
a hot medium. Indeed, as Fig. \ref{Fhot} demonstrates, the $D/\pi$
ratio should become significantly enhanced as compared to $pp$
collisions: assuming a fixed length $L=5\,\fm$ of the hot medium
traversed by the quarks, we find a factor of $\sim 2$ enhancement at
$p_{\perp} \sim 5 \div 10~\GeV$.

We only present the {\em ratio}\/ of quenching factors for the
following reason.
As noticed in \cite{BDMSquen}, in the region of transverse momenta we
are considering, the {\em absolute}\/ magnitude of quenching turns out
to be extremely sensitive to gluon radiation in the few-hundred-$\MeV$
energy range and therefore cannot be quantitatively predicted without
detailed understanding of the spectral properties of the medium.

The heavy-to-light ratio, however, proves to be much less sensitive to
the infrared region, since gluon radiation off heavy and light quarks
is universal in the $x\to 0$ limit, see \eqref{eq:spechq}. To
illustrate this point, in Fig.~\ref{Fhot} we show
the ratio of quenching factors calculated with the energy restriction
upon gluon energies, $\omega>500\,\MeV$.  We see that in the
$5-10\,\GeV$ range of $p_\perp$, this modifies the ratio by
$20-30$~\%.\footnote{The quenching factors themselves change
(increase) by an order of magnitude when the radiation of gluons with
energies smaller than $500\,\MeV$ is vetoed.}

The predicted ratio should not be taken at its face value: the
enhancement factor originates mainly from a very strong quenching of
light quark jets, and in reality such a dramatic suppression will be
washed out by light-particle production at the {\em periphery}\/ of
the collision zone.  In a central $AA$ collision, the number of
collisions in this peripheral zone, depending on the strength of
absorption, scales as $\sim A^{1/3} \div A^{2/3}$, whereas the total
number of collisions which contribute to the production of $D$ mesons
scales as $\sim A^{4/3}$. Thus, even with account of the collision
geometry, we may still expect an order of $\sim A^{2/3}
\div A$ enhancement of the $D/\pi$ ratio, which for a $Au-Au$ collision
may translate in a strong effect.

Moreover, given a large nucleus and moderately large $p_\perp$, both
light and (especially) heavy quarks will be tempted to turn into
hadrons {\em prior}\/ to leaving the interaction arena. In the pion
case, this is likely to add to quenching via pion absorption in the
medium~\cite{Kop}, thus enhancing further the $D/\pi$ ratio.

Clearly, detailed calculations have to be performed before a
defendable number for the magnitude of the $D/\pi$ enhancement can be
presented.  Nevertheless, the $D/\pi$ ratio appears to be extremely
sensitive to the density of colour charges in QCD matter, and we
eagerly await the results of experimental studies of this quantity in
relativistic heavy ion collisions.

%\vfill\eject
%\newpage

\section*{Acknowledgments}

We are grateful to Rolf Baier, Al Mueller and Dominique Schiff for
valuable discussions. D.K.\ wishes to thank the groups of LPTHE,
Universit\'e Pierre et Marie Curie and LPT, Universit\'e Paris-Sud,
for their hospitality during the period when this work was done.  The
research of D.K.\ is supported by the U.S.\ Department of Energy under
contract No.\ DE-AC02-98CH10886.

\end{document}